\documentclass[aps,prb,twocolumn,amssymb,floatfix,amsmath,aps,superscriptaddress,showpacs]{revtex4-1}
\usepackage{graphicx}
\usepackage{bm}
\usepackage[latin1]{inputenc}

\begin{document}

\title{Systematic motion of magnetic domain walls in notched nanowires under ultra-short current pulses}

\author{A. Pivano}
\affiliation{Aix-Marseille Universit\'{e}, CNRS, IM2NP UMR7334, F-13397 Marseille
    Cedex 20, France}

\author{V.O. Dolocan}
\email{voicu.dolocan@im2np.fr}
\affiliation{Aix-Marseille Universit\'{e}, CNRS, IM2NP UMR7334, F-13397 Marseille
    Cedex 20, France}

\begin{abstract}

The precise manipulation of transverse magnetic domain walls in finite/infinite nanowires with artificial defects under the influence of very short spin-polarized current pulses is investigated. We show that for a classical $3d$ ferromagnet material like Nickel, the exact positioning of the domain walls at room temperature is possible only for pulses with very short rise and fall time that move the domain wall reliably to nearest neighboring pinning position. The influence of the shape of the current pulse and of the transient effects on the phase diagram current-pulse length are discussed. We show that large transient effects appear even when $\alpha$=$\beta$, below a critical value, due to the domain wall distortion caused by the current pulse shape and the presence of the notches. The transient effects can oppose or amplify the spin-transfer torque (STT), depending on the ratio $\beta/\alpha$. This enlarges the physical comprehension of the DW motion under STT and opens the route to the DW displacement in both directions with unipolar currents.

\end{abstract}

\date{\today}
\maketitle

%%%%%%%%%%%%%%%%%%%%%%%%%%%%%%%%%%%%%%%%%%%%%%%%%%%%%%%%%%%%%%%%%%%%%%%%%%%%%%%%%%%%%%%%%%%%%%%%%%%%%%%%%%%%
%%%%%%%%%%%%%%%%%%%%%%%% SECTION Introduction %%%%%%%%%%%%%%%%%%%%%%%%%%%%%%%%%%%%%%%%%%%%%%%%%%%%%%%%%%%%%%%%%%%%%%%%%%
%%%%%%%%%%%%%%%%%%%%%%%%%%%%%%%%%%%%%%%%%%%%%%%%%%%%%%%%%%%%%%%%%%%%%%%%%%%%%%%%%%%%%%%%%%%%%%%%%%%%%%%%%%%%

\section{Introduction}\label{Introduction}

Current induced magnetic domain wall motion (CIDWM) in nanowires or nanostrips is a highly active research field\cite{Thomasbook,Boulle} with applications in high-density and ultrafast nonvolatile data-storage devices like the racetrack memory\cite{Parkin} or for logic devices\cite{Allwood}. In the racetrack memory, the data processing is based on the controlled displacement between precise distinct positions of the domain walls (DWs) due to the transfer of angular momentum (spin-transfer torque) from a spin-polarized electric current. To achieve precise positioning of DW, artificial constrictions or others patterned geometrical traps are usually used, which create an attractive pinning potential for the DW. Different types of traps were studied in cylindrical or flat/strip nanowires\cite{Petit1, Petit2, Martinez2009, Dolocan1}, along with the possible interaction between the DWs\cite{Dolocan2, Pivano1}. In some cases, depending on the pinning potential, the DW displacement between pinning sites can display a chaotic behavior\cite{Pivano2} or a stochastic resonance\cite{Martinez2011} under harmonic excitation.

The required currents for STT based DW movement are usually high ($\sim$ 1A/$\mu$m$^2$) which limits the applicability due to Joule heating. To displace accurately the DWs between the pinning sites, the current density should be kept at relatively low values and/or very short current pulses should be applied. Experimentally, it was observed that an efficient DW motion is reached for pulses in the nanosecond regime\cite{Thomas1} and that the resonant excitation of the DW by a short train of current pulses decreases the depinning current\cite{Thomas2}. More recently, the effect of the temporal and spatial shape of the current pulse was highlighted\cite{Bocklage, Yan, Langner}. It was shown that a fast changing current with an ultra-short pulse rise time decreases the critical current density due to the dependence of the DW motion on the time derivative of the current\cite{Kruger} and leads to high DW velocity\cite{Heyne}. Another aspect of the CIDWM under short pulses is the existence of important transient effects related with the DW inertia-like behavior\cite{Thomas1,Thiaville1,Chauleau,Thomas3,Rhensius} due to deformation of the wall. The consequences are a delayed response at the current onset and at the end of the current pulse. The theoretical and experimental results show that the distance traveled by a DW is almost proportional to the current pulse length and that the transient motion depends on the variation of the generalized angle of the wall, the wall width and the ratio of the damping ($\alpha$) and nonadiabatic ($\beta$) parameters\cite{Thiaville1,Chauleau}. For very short pulses (one nanosecond), the transient displacement is comparable with the steady-state motion\cite{Taniguchi}. A DW that propagates without deformation should display no inertia\citep{Wang} like in cylindrical nanowires\cite{Hertel} or in certain perpendicular magnetic anisotropy systems\cite{Vogel}. The absence of inertia will allow a fast response to external forces while the transient DW displacement after the application of the pulse limits the application in fast devices. In the main time, it was recently demonstrated that inertia-like behavior of a DW can be also an advantage when ultra-short optical pulses are used\cite{Janda} with applications in the optical recording. Moreover, in systems with strong spin-orbit coupling where additional contributions from spin-Hall effect complicate the DW dynamics, a DW tunable inertia was proposed\cite{Torrejon}. Before studying more complex systems, the influence of the transient effects on the systematic DW movement under ultra-short spin-polarized current pulses should be completely understood in a classical $3d$ ferromagnet like Nickel, which is the aim of this paper. 

In this paper, we address the systematic motion of a magnetic transverse DW between fixed artificial constrictions (notches), when submitted to a series of ultra-short spin-polarized current pulses (transient regime) at low and room temperature. The artificial constrictions are situated at regular positions in a flat finite or infinite nanowire with only in-plane (shape) anisotropy like in a classical $3d$ ferromagnet.  We determine the influence of the current pulse shape (rise and fall time) on the motion of the DW. We show that even at zero temperature, there is a transition region between the different bands in the current-pulse time phase diagram, each band corresponding to the positioning of the DW at a well-defined notch. Our results show that at room temperature, the precise positioning can be achieved only by very short pulses with very short rise and fall times that displace the DW by only one notch at a time. Therefore, to move a DW several notches reliably, a sequence of very short pulses should be used. By examining the influence of the damping and nonadiabatic parameters, we show that when $\beta$=0 the transient effects (automotion) of the DW are very large and oppose the STT, being observed in the phase diagrams as predicted\cite{Thiaville1}. The transient effects are related with the change in the DW structure that is due to a combination of factors: pinning potential of the notches which induces a sufficient variation of the DW angle\cite{Yuan2014}, position of the DW inside the potential well (different restoring force), low damping parameter and shape of the current pulse. Contrary to expectations, the transient effects also appear when $\alpha=\beta$, below a critical value. For $\beta > \alpha$, the transient effects can oppose or amplify the STT, thus explaining the oscillatory DW depinning at higher currents observed experimentally\cite{Thomas1}.  This brings new physical insight into CIDWM under STT and paves the way for systematically displacing DW in nanowires in both directions using only unipolar current pulses. 

This article is organized as follows. In Sec.~\ref{Model}, we present the micromagnetic and the stochastic 1D model used to calculate the pulsed DW dynamics. In Sec.~\ref{Results}, we compute and investigate the phase diagram of the DW dynamics for a finite and infinite nanostrip at T=0K and room temperature. Discussion and concluding remarks are presented in Sec.~\ref{Discussion}.

%%%%%%%%%%%%%%%%%%%%%%%%%%%%%%%%%%%%%%%%%%%%%%%%%%%%%%%%%%%%%%%%%%%%%%%%%%%%%%%%%%%%%%%%%%%%%%%%%%%%%%%%%%%%
%%%%%%%%%%%%%%%%%%%%%%%%%%  FIGURE 1   %%%%%%%%%%%%%%%%%%%%%%%%%%%%%%%%%%%%%%%%%%%%%%%%%%%%%%%%%%%%%%%%%%%%%%%%%%%%%%%%%%
%%%%%%%%%%%%%%%%%%%%%%%%%%%%%%%%%%%%%%%%%%%%%%%%%%%%%%%%%%%%%%%%%%%%%%%%%%%%%%%%%%%%%%%%%%%%%%%%%%%%%%%%%%%%

\begin{figure*}[ht]
  \includegraphics[width=17cm]{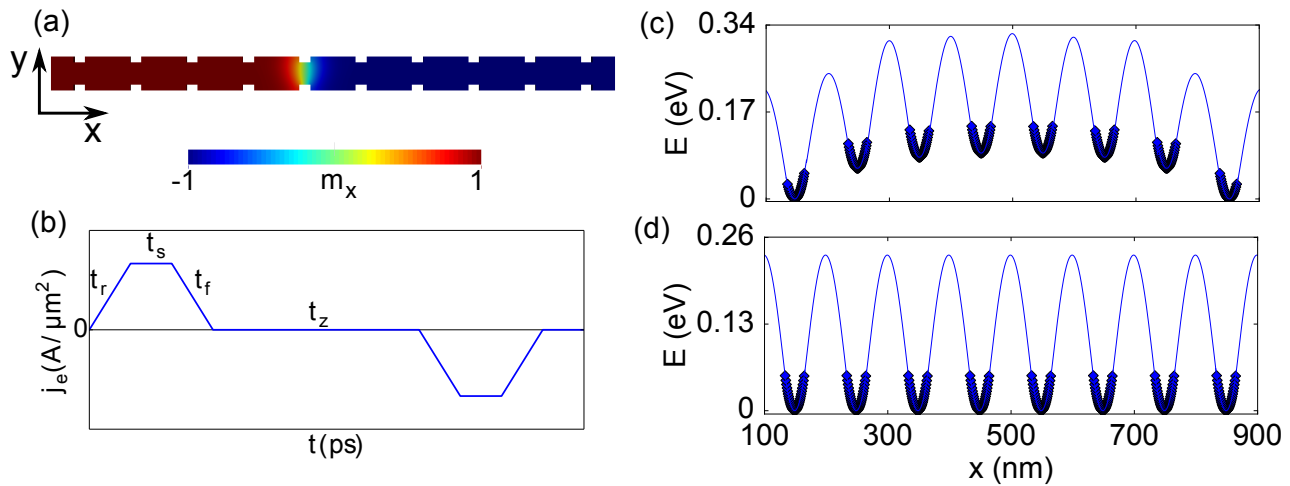}\\
 \caption{\label{Fig.1} (Color online) (a) Simulated structure: planar nanowire with ten symmetric double notches. The equilibrium position of a pinned DW is shown in the finite case. (b) Definition of current pulse with its temporal characteristics. Two successive current pulses with opposite polarity are shown. The normalized potential pinning energy for finite and infinite case as determined by micromagnetic simulations (symbols) are shown in (c) and (d) respectively. The line is a fit as described in the text. }
\end{figure*}

%%%%%%%%%%%%%%%%%%%%%%%%%%%%%%%%%%%%%%%%%%%%%%%%%%%%%%%%%%%%%%%%%%%%%%%%%%%%%%%%%%%%%%%%%%%%%%%%%%%%%%%%%%%%
%%%%%%%%%%%%%%%%%%%%%%%%%%%%%%%%%%%%%%%%%%%%%%%%%%%%%%%%%%%%%%%%%%%%%%%%%%%%%%%%%%%%%%%%%%%%%%%%%%%%%%%%%%
%%%%%%%%%%%%%%%%%%%%%%%%%%%%%%%%%%%%%%%%%%%%%%%%%%%%%%%%%%%%%%%%%%%%%%%%%%%%%%%%%%%%%%%%%%%%%%%%%%%%%%%%%%%%

%%%%%%%%%%%%%%%%%%%%%%%%%%%%%%%%%%%%%%%%%%%%%%%%%%%%%%%%%%%%%%%%%%%%%%%%%%%%%%%%%%%%%%%%%%%%%%%%%%%%%%%%%%%%
%%%%%%%%%%%%%%%%%%%%%%%% SECTION MODEL %%%%%%%%%%%%%%%%%%%%%%%%%%%%%%%%%%%%%%%%%%%%%%%%%%%%%%%%%%%%%%%%%%%%%%%%%%%%%%%%%%%%
%%%%%%%%%%%%%%%%%%%%%%%%%%%%%%%%%%%%%%%%%%%%%%%%%%%%%%%%%%%%%%%%%%%%%%%%%%%%%%%%%%%%%%%%%%%%%%%%%%%%%%%%%%%%

\section{Model}\label{Model}

We study numerically the systematic motion of a pinned transverse domain wall in a finite or infinite Ni nanostrip with symmetric rectangular notches. The finite strip has a length L$_x$=1$\mu$m, a cross section of L$_y \times$L$_z$ = 60$\times$5nm$^2$ and has ten rectangular symmetric double notches separated by 80nm. The results presented below are for notch dimensions of 20 $\times$ 9 $\times$ 5 nm$^3$. The variation of length and depth of the notches does not influence much the physics of phase diagrams presented in Section~\ref{Results}. The notch depth influences the depinning current as the potential barrier increases, while the notch length influences lightly the depinning current and mostly the slope of the potential wells.

Fig.~\ref{Fig.1}(a) shows the equilibrium position of a head-to-head transverse DW in the notched nanostrip using the parameters of Nickel: saturation magnetization M$_s$=477kA/m, exchange stiffness parameter A = 1.05 $\times$ 10$^{-11}$J/m, spin polarization P=0.7 and damping parameter $\alpha$=0.05.  The DW is moved by a series of spin polarized current pulses applied along the $x$-axis. The geometry of the current pulse is described in Fig.~\ref{Fig.1}(b), with t$_r$, t$_s$, t$_f$ and t$_z$  the rise, stable, fall time and zero-current time respectively.  The nonadiabatic parameter is set to $\beta$=2$\alpha$, if not specified otherwise.

The DW dynamics was computed using 3D micromagnetic simulations with the M{\small U}M{\small AX}3 package \cite{ArneMumax} and with the one-dimensional DW model\cite{Slonc,Thiaville}. In both cases, the magnetization dynamics is determined from the Landau-Lifschitz-Gilbert (LLG) equation with adiabatic and non-adiabatic spin-transfer torques\cite{ZhangLi}:

%%%%%%%%%%%%%%%%%%%%%%%%%%%%%%%%%%%%%%%%%LLG equation%%%%%%%%%%%%%%%%%%%%%%%%%%%%%%%%%%%%%%%%%%%%%%%%%%%%%%%%%%%%%%%%%%%

\begin{equation}\label{LLGST}
\mathbf{\dot{M}} =   - \gamma_0 \mathbf{M} \times \mathbf{H}_{eff} + \alpha ( \mathbf{M} \times \mathbf{\dot{M}}) - ( \mathbf{u} \cdot\nabla)\mathbf{M} +  \beta\mathbf{M}\times(\mathbf{u}\cdot\nabla)\mathbf{M}
\end{equation}

%%%%%%%%%%%%%%%%%%%%%%%%%%%%%%%%%%%%%%%%%%%%%%%%%%%%%%%%%%%%%%%%%%%%%%%%%%%%%%%%%%%%%%%%%%%%%%%%%%%%%%%%%%%%

\noindent where $\gamma_0$ is the gyromagnetic ratio, $\mathbf{u} = \mathbf{j}_e P \mu_B / eM_s$ is the spin drift velocity, $P$ the spin polarization of conductions electrons, $\mu_B$ the Bohr magneton and $\mathbf{j}_e$ the applied current density. No additional exotic torques (like the ones due to the spin-Hall or Rashba effect) were considered.

For the micromagnetic computations, the strip was discretized into a mesh with a cell size of 2$\times$3$\times$2.5nm$^{3}$, inferior to the exchange length ($\sim$5nm). The DW dynamics in a finite wire is compared with the one of an infinity long wire where the magnetic charges at the ends of the nanostrip are compensated. The average position of the DW center ($X$) is extracted for each simulation (in the axial $x$ direction) along with the azimuthal angle ($\psi$) of magnetization in the yz plane. No magnetocrystalline anisotropy is considered, the shape anisotropy insures that the easy axis is in-plane. The effect of the temperature is studied both micromagnetically and with the 1D model. The 1D model of the DW (collective coordinates $X$ and $\psi$) supposes that the DW is rigid and gives a quasi-quantitative understanding of the motion of the DW. The Langevin equations of motion of the DW\cite{Boulle,Lucassen} are detailed in Ref.~\onlinecite{SM}.

The pinning potential energy is determined from quasistatic micromagnetic simulations and is shown in Fig.~\ref{Fig.1}(c) and (d) for the finite case and the infinite case respectively. The pinning potential determined by fitting the micromagnetic results is harmonic inside the notches and sinusoidal between them:

%%%%%%%%%%%%%%%%%%%%%%%%%%%%%%%%%%%%%%%%%Potential equation%%%%%%%%%%%%%%%%%%%%%%%%%%%%%%%%%%%%%%%%%%%%%%%%%%%%%%%%%%%%%%%%%%%
\begin{equation}
\label{PinningPotential}
V_{p}(x)=
\begin{cases}
\frac{1}{2} k_{i} (x+x_i)^{2}, & \text{for }  x_i - L \leq x \leq x_i + L \\
V_{0} \cos (2\pi f x + \phi_{i,i+1}), & \text{for } x_{i} + L < x \leq x_{i+1}- L
\end{cases}
\end{equation}
%%%%%%%%%%%%%%%%%%%%%%%%%%%%%%%%%%%%%%%%%%%%%%%%%%%%%%%%%%%%%%%%%%%%%%%%%%%%%%%%%%%%%%%%%%%%%%%%%%%%%%%%%%%%

\noindent with $k_{i}$ the stiffness constant and $x_{i}$ the DW stable equilibrium position of the site $i$, $\phi_{i,i+1}$ the phase between the $i$ site and its nearest neighbor. $V_{0}$ and $f$ correspond respectively to the effective height of the potential and its spatial frequency. For the finite case, the stiffness constant varies from 7.07$\times 10^{-5}$ N/m to 6.77$\times 10^{-5}$ N/m, when moving from the center of the nanowire to its ends. In the infinite case, the stiffness constant is equal to 7.16 $\times 10^{-5}$ N/m for all the pinning sites and $L$ = 16.5nm. The expression (Eq.~\ref{PinningPotential}) was used in the equations of the 1D model through the pinning field $H_{p}$ included in $H_{eff}$.

The pinning energy is controlled by the dimensions and distance between the notches. A clear difference is observed between the two potentials due to edge dipolar energy. In the finite strip, the depinning field decreases from 39 Oe, in the central wells, to 26 Oe when the DW is closed to the two ends of the strip. This is due to the attractive interaction between the DW and magnetic surface charges located at the sides\cite{Kruger2}. As a result, the potential wells are asymmetric in energy along the strip and their energy minima decrease when the distance between the notches and the ends of the strip is reduced. In contrast, for the infinite case, each well has the same depinning field and energy barriers. 

%%%%%%%%%%%%%%%%%%%%%%%%%%%%%%%%%%%%%%%%%%%%%%%%%%%%%%%%%%%%%%%%%%%%%%%%%%%%%%%%%%%%%%%%%%%%%%%%%%%%%%%%%%%%
%%%%%%%%%%%%%%%%%%%%%%%% SECTION results %%%%%%%%%%%%%%%%%%%%%%%%%%%%%%%%%%%%%%%%%%%%%%%%%%%%%%%%%%%%%%%%%%%%%%%%%%%%%%%%%%
%%%%%%%%%%%%%%%%%%%%%%%%%%%%%%%%%%%%%%%%%%%%%%%%%%%%%%%%%%%%%%%%%%%%%%%%%%%%%%%%%%%%%%%%%%%%%%%%%%%%%%%%%%%%

\section{Results}\label{Results}

Our analysis of the DW dynamics begins with the study of the differences between a finite and an infinite nanostrip at T=0K. Afterwards, the influence of the pulse shape is discussed and the particularities of the DW motion at room temperature. The last subsection details the results when the damping and nonadiabatic parameters are varied and their influence on the transient displacement. 

\subsection{Phase diagrams at T = 0K}\label{subsec:0K}

%%%%%%%%%%%%%%%%%%%%%%%%%%%%%%%%%%%%%%%%%%%% FIGURE 2 %%%%%%%%%%%%%%%%%%%%%%%%%%%%%%%%%%%%%%%%%%%%%%%%%%%%%%%%%%%%%%%%
%%%%%%%%%%%%%%%%%%%%%%%%%%%%%%%%%%%%%%%%%%%%%%%%%%%%%%%%%%%%%%%%%%%%%%%%%%%%%%%%%%%%%%%%%%%%%%%%%%%%%%%%%%%%
\begin{figure}[ht]
  \includegraphics[width=\linewidth]{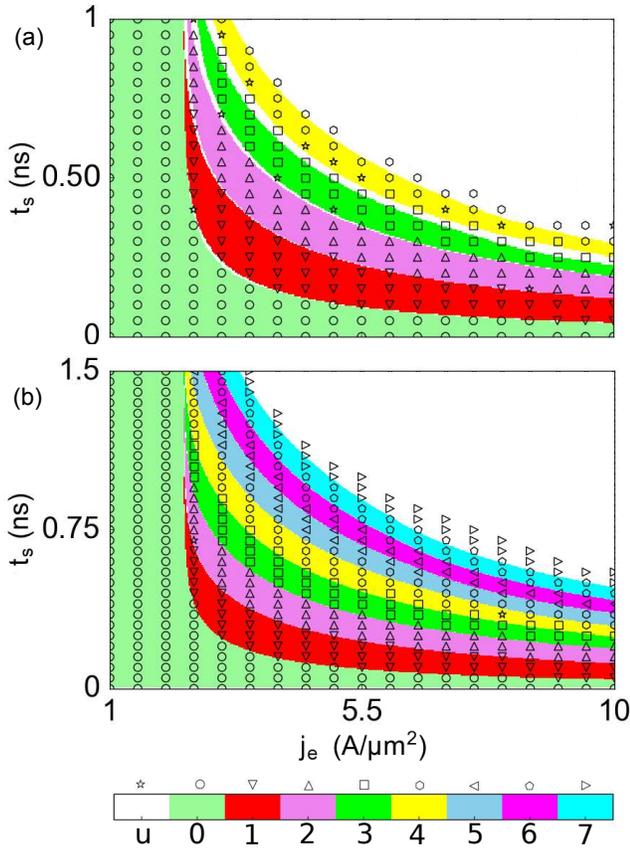}\\
 \caption{\label{Fig.2} (Color online) Phase diagram for a DW at T=0K in a finite (a) and infinite (b) nanostrip in the parameter space stable-time -- current amplitude with t$_r$=t$_f$=5ps and t$_z$=10ns. The total time of the periodic pulses is 350ns. The micromagnetic results (scattered symbols) are compared to the 1D model (colored regions). The diagrams show only a few bands due to the finite size of the wire or due the number of notches used. The upper right region is due to the finite size of the nanowire or of the simulated window (infinite case).}
\end{figure}
%%%%%%%%%%%%%%%%%%%%%%%%%%%%%%%%%%%%%%%%%%%%%%%%%%%%%%%%%%%%%%%%%%%%%%%%%%%%%%%%%%%%%%%%%%%%%%%%%%%%%%%%%%%%
%%%%%%%%%%%%%%%%%%%%%%%%%%%%%%%%%%%%%%%%%%%%%%%%%%%%%%%%%%%%%%%%%%%%%%%%%%%%%%%%%%%%%%%%%%%%%%%%%%%%%%%%%%%%

To characterize the systematic motion of the DW between the notches, we computed point-by-point phase diagrams for all systems with the 1D DW model for a large range of pulse duration and current amplitude. We compare our 1D results with phase diagrams computed micromagnetically on less points than the 1D calculus. A similar micromagnetic computation will require an enormous execution time. The control parameters are the amplitude, the duration and the shape of the current pulses. The range of the current amplitude ($\leq$10 A/$\mu$m$^{2}$) is chosen to have only viscous motion (no precession) for the pulse duration used ($\lesssim$ 1.5ns). The pulse duration range is selected to be on the same order of magnitude with access or reading/writing time in possible magnetic memories based on DW. 

The phase diagram, in the parameter space stable time -- current amplitude, which characterizes the DW dynamics in the finite nanostrip at T=0K, is shown in Fig.~\ref{Fig.2}(a). The diagram represents 200$\times$400 point-by-point integration with a fourth order Runge-Kutta scheme. The micromagnetic results (scattered symbols) are compared with the 1D results (colored regions). The DW is initialized in the left central well and a series of periodic bipolar current pulses are applied during 350ns to move periodically forth and back the DW between two desired notches. The pulse characteristics are t$_r$=t$_f$=5ps and t$_z$=10ns. The influence of t$_r$ and t$_f$ is discussed below. The value of t$_z$ (10ns) was chosen to correspond to the return to equilibrium time of the DW at room temperature. The t$_z$ can be reduced to 3ns for T=0K, without a change of the phase diagrams. In Fig.~\ref{Fig.2}(a), several regions are visible, each region corresponding to one state of the DW oscillation. The first state which appears is the pinned state, noted state 0 in the micromagnetic simulation, and corresponds to the DW being pinned in the initial notch. After the pulse ends, the DW behaves like a damped harmonic oscillator. The state 0 is observed for all t$_s$, when the external current j$_e$ is inferior to the depinning current 2.31 A/$\mu$m$^{2}$. This state is also observed at higher currents (until 10 A/$\mu$m$^{2}$), when the stable time t$_s$ is low (between 0ps and 55ps). The diagram displays others bands, where the DW oscillates periodically between the same two potential wells, which can be next-neighbors (noted as band 1) or not, until the fifth state that corresponds to the periodic oscillation between the initial notch and the fourth notch to the right (band 4) hoping the three notches in between. The number of bands is given by the considered finite size of the nanostrip. The second state (next-neighbors notches noted band 1) is observed up to t$_s$=0.96ns at j$_e$=2.3308 A/$\mu$m$^{2}$, while the others bands continue above t$_s$=1ns. Thus, the DW can cross severals notches back and forth for a given current pulse characteristic. We observe that the interband transitions are characterized by an unintended state (state u), where the DW oscillation does not take place between the desired positions. This transition is more pronounced between the last two bands. The micromagnetic results, which are superimposed on the 1D results, give quantitatively the same results until the fourth state, after which a small shift appears in the t$_s$ and j$_e$ values, but the bands are qualitatively the same. The upper right region, which correspond to an unwanted state, is due to the finite size of the nanowire, here the DW is pinned at the nanowire end.

In the infinite case, the DW is initialized in the first well from the left. A phase diagram similar to the finite case is shown in Fig.~\ref{Fig.2}(b). This diagram contains three more bands than the finite case, which correspond to additional states where the DW oscillates between two notches, starting from the initial one, separated by four intermediate notches (band 5) until six intermediate notches (band 7). To have a better visibility over these new bands we computed the DW dynamics for t$_s$ up to 1.5ns on 300$\times$400 points. We observe that the band 1 and 2 exist until t$_s$=1.045ns and t$_s$=1.27ns respectively at j$_e$=2.3308 A/$\mu$m$^{2}$, while the other bands continue above 1.5ns. The interband transition (band u) is observed between the bands zero and one around j$_e$=3 A/$\mu$m$^{2}$ and t$_s$=250ps. For the superior bands, the interband transition at boundaries is quasi-nonexistent, which shows that the infinite case is more stable than the finite case. In both cases, the 1D model gives quantitatively the same results as the micromagnetic simulations in the three first bands and quasi-quantitatively in the others (a shift in values is visible). As for the finite case, the upper right region appears due to the finiteness of the simulated length of the nanowire, even if the end charges are suppressed. If a longer simulated window is considered, other superior bands will follow as seen for example in Fig.~\ref{Fig.5} or Fig.~\ref{Fig.6}.

%%%%%%%%%%%%%%%%%%%%%%%%%%%%%%%%%%%%%%%%%%%%%%%%%%%%%%%%%%%%%%%%%%%%%%%%%%%%%%%%%%%%%%%%%%%%%%%%%%%%%%%%%%%%
\begin{figure}
  \includegraphics[width=\linewidth]{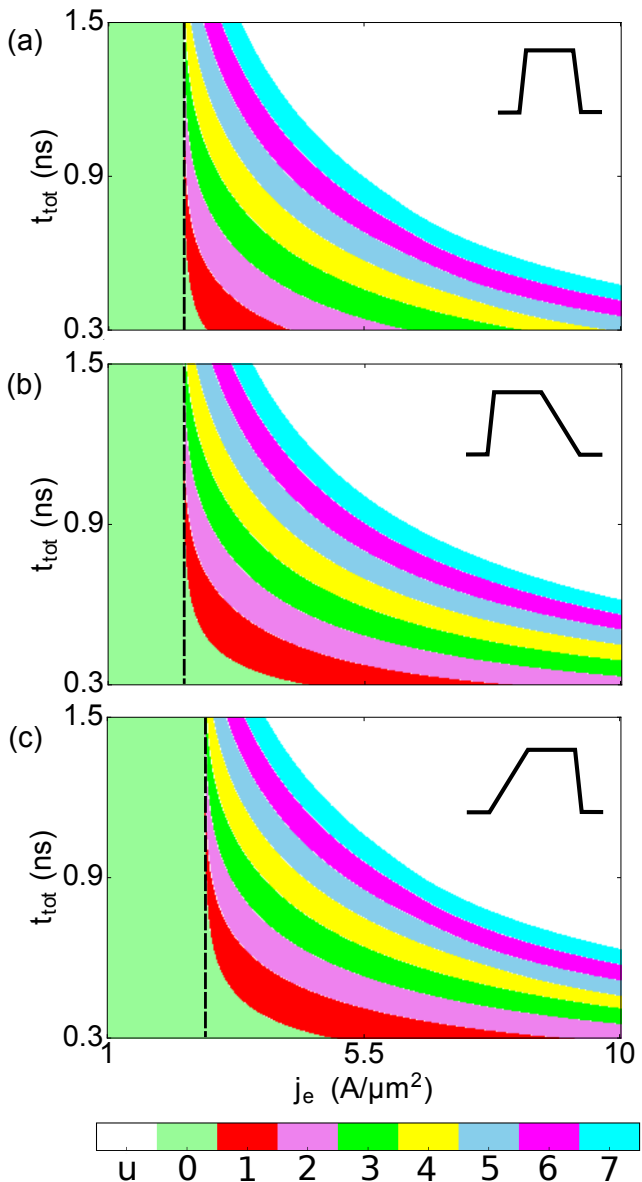}\\
 \caption{\label{Fig.3} (Color online) Influence of the pulse shape on the phase diagram for a DW at T=0K in an infinite nanostrip. The parameter space is the total time (t$_r$ + t$_s$+ t$_f$)  vs. current amplitude. (a) Symmetric pulse with t$_r$ = t$_f$ = 5 ps and t$_z$ = 10 ns, Asymmetric pulse with (b) t$_r$ = 5 ps, t$_f$ = 300 ps and t$_z$ = 10 ns, and (c) t$_r$ = 300 ps, t$_f$ = 5 ps and t$_z$ = 10 ns. The upper right region is due to the finite size of the simulated window.}
\end{figure}
%%%%%%%%%%%%%%%%%%%%%%%%%%%%%%%%%%%%%%%%%%%%%%%%%%%%%%%%%%%%%%%%%%%%%%%%%%%%%%%%%%%%%%%%%%%%%%%%%%%%%%%%%%%%

The influence of the pulse shape is detailed in Fig.~\ref{Fig.3} for the infinite wire as calculated with the 1D model. In panel (a), the pulse shape is symmetric as t$_r$ = t$_f$ = 5ps and t$_z$ = 10ns, while in panel (b) and (c) the pulse shape is asymmetric, t$_r \neq$ t$_f$ with t$_z$ kept constant at 10ns. The values of t$_r$ and t$_f$ were varied between 5ps and 300ps. In panel (b), the case with  t$_r$ = 5ps and t$_f$ = 300ps is shown, while in panel (c), the one with t$_r$ = 300ps and t$_f$ = 5ps. In the three panels, the total time (t$_{tot}$ = t$_r$ + t$_s$+ t$_f$) is shown starting from 300ps, to be able to compare the diagrams evolution with the pulse shape. We observe that the first depinning current depends mainly on the rise time as depicted in panels (a) and (b), where only the fall time is varied. In these cases, the first depinning current is the same and equal to 2.31 A/$\mu$m$^{2}$. The influence of the t$_f$ is an offset of the bands along the total time axis, therefore if t$_f$ is decreased the second band is shifted to shorter times and almost disappears from the shown phase diagram. The influence of the rise time manifests itself also as an offset of the bands to larger times, but also to larger currents, therefore a higher first depinning current equal to 2.69 A/$\mu$m$^{2}$. The first depinning current for a total time of 0.3ns is 2.78 A/$\mu$m$^{2}$ for the symmetric pulse, raising to 4.54 A/$\mu$m$^{2}$ (panel (b)) and 5.03 A/$\mu$m$^{2}$ (panel (c)) in the asymmetric case. The micromagnetic calculation (not shown) gives similar results as the ones shown in Fig.~\ref{Fig.2}, meaning a small offset of the bands compared with the ones calculated with the 1D model starting from the fourth band.

The dependence of the depinning current on the rise time was deduced from the linearized equation of motion in the 1D approximation, as the force on the DW can be written as\cite{Kruger,Bocklage}  :

\begin{equation}\label{eq1}
\ddot{X} = - \frac{\dot{X}}{\tau_d} - \frac{1}{m} \frac{dE}{dX} + \frac{\beta}{\alpha \tau_d} u + \frac{1+\alpha\beta}{1+\alpha^2} \dot{u}
\end{equation}

\noindent where $m = \frac{2\alpha S \mu_0 M_s \tau_d}{\Delta \gamma_0}$ the DW mass, $\tau_d = \frac{1+\alpha^2}{\alpha\gamma_0 H_k}$ is the damping time, with $H_k$ the anisotropy field, $\Delta$ the DW width and $E$ the pinning energy. The force on the wall depends on the current and its derivative, therefore a shorter rise time increases the derivative term which leads to a decreasing of the depinning current and vice versa. For the present results, the damping time is 0.27ns (0.68ns for $\alpha = 0.02$) therefore the DW is in the transient regime for the pulse duration used.

%%%%%%%%%%%%%%%%%%%%%%%%%%%%%%%%%%%%%%%%%%%%%%%%%%%%%%%%%%%%%%%%%%%%%%%%%%%%%%%%%%%%%%%%%%%%%%%%%%%%%%%%%%%%
%%%%%%%%%%%%%%%%%%%%%%%%%%%%%%%%%%%%%%%%%%%%%%%%%%%%%%%%%%%%%%%%%%%%%%%%%%%%%%%%%%%%%%%%%%%%%%%%%%%%%%%%%%%%
\subsection{Temperature dependence}\label{subsec:temp}
%%%%%%%%%%%%%%%%%%%%%%%%%%%%%%%%%%%%%%%%%%%%%%%%%%%%%%%%%%%%%%%%%%%%%%%%%%%%%%%%%%%%%%%%%%%%%%%%%%%%%%%%%%%%
%%%%%%%%%%%%%%%%%%%%%%%%%%%%%%%%%%%%%%%%%%%%%%%%%%%%%%%%%%%%%%%%%%%%%%%%%%%%%%%%%%%%%%%%%%%%%%%%%%%%%%%%%%%%

%%%%%%%%%%%%%%%%%%%%%%%%%%%%%%%%%%%%%%%%%%%%%%%%%%%%%%%%%%%%%%%%%%%%%%%%%%%%%%%%%%%%%%%%%%%%%%%%%%%%%%%%%%%%
\begin{figure*}[ht]
  \includegraphics[width=18cm]{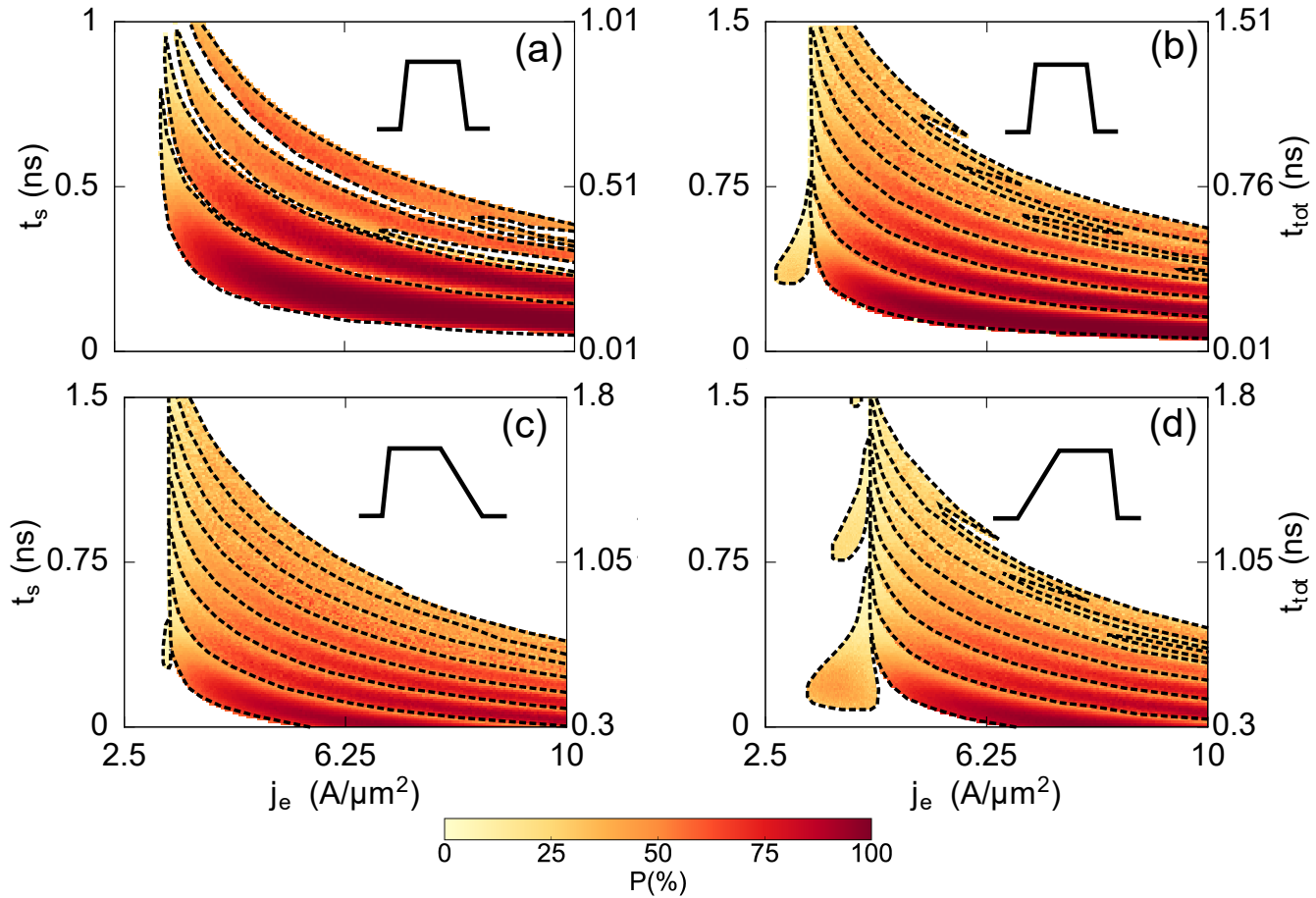}\\
 \caption{\label{Fig.4} (Color online) Probability of DW motion in different bands for the finite strip (a) and the infinite strip (b)-(d) at T=293K. (a) and (b) A symmetric pulse is applied with the characteristics t$_r$ = t$_f$ = 5 ps, after an initial and final t$_z$ of 10 ns. (c) An asymmetric pulse is applied with pulse shape t$_r$ = 5 ps, t$_f$ = 300 ps, after an initial and final t$_z$ of 10 ns. (d) An asymmetric pulse is applied with pulse shape t$_r$ = 300 ps, t$_f$ = 5 ps, after an initial and final t$_z$ of 10 ns. The damping parameter is taken as $\alpha = 0.02$ and $\beta=2\alpha$. The band pockets, which appear in the panels (b) to (d) on the left, correspond to the band -1. The dotted lines are guide to the eyes and represent the bands edges.}
\end{figure*}
%%%%%%%%%%%%%%%%%%%%%%%%%%%%%%%%%%%%%%%%%%%%%%%%%%%%%%%%%%%%%%%%%%%%%%%%%%%%%%%%%%%%%%%%%%%%%%%%%%%%%%%%%%%%

The temperature influences the systematic motion of the DW by modifying the DW relaxation in a potential well after an applied current pulse. The oscillations during the DW relaxation could be sustained by the thermal noise, which could lead to a jump to the wrong well while the following pulse occurs, or on the contrary, the thermal noise could counter the effect of the current pulse and the DW could stay pinned in the non-desired potential well.

To carry out this study, we computed the DW dynamics at T=293K for finite/infinite nanostrip and different pulse shapes. The results are shown in Fig.~\ref{Fig.4}. In all cases, panels (a) to (d), the DW first oscillates freely (relaxation) in the presence of the thermal noise in its initial well during 10ns and afterwards a current pulse is applied to push the DW to another well (corresponding to one of the bands in Figs.~\ref{Fig.2} and \ref{Fig.3}), followed by a DW relaxation during another 10ns. The damping parameter $\alpha$ is taken as 0.02, lower than the one at T=0K\cite{Gilmore}, and the non-adiabatic parameter is taken as $\beta$ = 2$\alpha$ = 0.04. The same phase diagrams, as presented in Figs.~\ref{Fig.2} and \ref{Fig.3}, were recalculated with $\alpha$ = 0.02 and $\beta$ = 0.04 at T=0K and these bands, from one to four, are indicated by dotted lines in Fig.~\ref{Fig.4}(a) for the finite nanostrip, while the bands from one to seven are indicated in panels (b) to (d) for the infinite nanostrip. Starting from the fourth band, the shape of the bands changes showing a reentrant transition (except panel (c)), and the phase diagrams from panels (b) to (d) show band pockets to the left corresponding to negative DW displacement of one notch (noted as band -1) even though the STT pushes the DW in the positive direction. These features are discussed in Sec.~\ref{subsec:automotion}.

The stochastic motion of the DW was computed for a number of bands with the stochastic 1D model, and only on a certain number of points micromagnetically. The Fig.~\ref{Fig.4}(a) shows the probability of the DW motion in the first four bands for a finite nanowire, when a symmetric current pulse (t$_r$ = t$_f$ = 5ps) is applied after a relaxation time of 10ns. At the end of the pulse, the DW is relaxed another 10ns before its position is considered acquired. A certain number of realizations was computed for a quarter of the phase diagram points of each band: 2700 realizations for each point shown from the first band and 500 realizations for each point of the superior bands. The maximum of probability (100$\%$) for the precise positioning of the DW to the nearest notch is found only for 0.76$\%$ of the first band's calculated points (17 points), while on 32.18$\%$ of the points the probability is superior to 95$\%$. The maximum of probability decreases rapidly with increasing the band number, being 98.6$\%$ (for 3 points) for the second band, 67.4$\%$ for the third band and 71$\%$ for the fourth band. These results are to be compared with panel (b), where the same pulse is applied in an infinite nanowire and the same number of realizations were computed for each band. The maximum probabilities are similar for the first two bands, for similar band point number density, indicating that at room temperature only few current pulse characteristics give 100$\%$ probability of precise positioning. The points in the first band, that correspond to 100$\%$ probability of desired motion, appear for an applied current superior to 7.7  A/$\mu$m$^{2}$ and a pulse length between 100ps and 130 ps for the finite strip and superior to 8.1 A/$\mu$m$^{2}$ and a pulse length between  90ps and 120 ps for the infinite case respectively. For superior bands, starting with the fourth, the maximum of probability and band point number density are increased in the infinite nanostrip compared to the finite case, as in the latter the potential barrier is weaker for the more distant notches (see Fig.~\ref{Fig.1}(c)). Micromagnetically, we computed the probability for the phase diagram points corresponding to the 100$\%$ values found with the 1D model (which appear only in the first band) on 200 realizations/point. These probabilities vary between 92$\%$ and 97$\%$. The small difference between the probability calculations of the micromagnetic and 1D model is attributed to the small shift in the phase diagrams that was shown to exist between the two.

Fig.~\ref{Fig.4}(c) and (d) show the probability of DW motion when an asymmetric current pulse is applied after 10ns initial and final relaxation. The pulse shape is t$_r$ = 5ps, t$_f$ = 300ps in panel (c) and t$_r$ = 300ps, t$_f$ = 5ps in panel (d). The maximum of probability in the first band of panel (c) is 96.77$\%$, with only 2.31$\%$ of calculated band points having a probability superior to 95$\%$ (from 2700 realizations). For the superiors bands, the maximum of probability diminishes with only 2.93$\%$ of points having a probability superior to 90$\%$ in the second band and 2.32$\%$ of points having a probability superior to 80$\%$ in the third band. For panel (d), the maximum of probability in the first band is found to be 98.07$\%$, with 7.69$\%$ of points having a probability superior to 95$\%$. The maximum of probability diminishes faster in the superior bands as compared with the panel(c) results, with less points having maximum bands probability. For example, only 0.79$\%$ of points have a probability superior to 90$\%$ in the second band and 0.54$\%$ of points have a probability superior to 80$\%$ in the third band. The difference in probability between the two asymmetric pulses is due to the fact that the long rise time of the pulse (panel (d)) produces the shift up and right of the bands compared with panel (c), therefore more points with maximum probability are found in the first band (as these points are usually close to the band center) and lower points with maximum probability in the superior bands as there are less points on these bands.  

These results suggest that to achieve a well defined DW positioning at room temperature by STT alone, individual pulses should be applied with very short rise and fall time.

%%%%%%%%%%%%%%%%%%%%%%%%%%%%%%%%%%%%%%%%%%%%%%%%%%%%%%%%%%%%%%%%%%%%%%%%%%%%%%%%%%%%%%%%%%%%%%%%%%%%%%%%%%%%
%%%%%%%%%%%%%%%%%%%%%%%%%%%%%%%%%%%%%%%%%%%%%%%%%%%%%%%%%%%%%%%%%%%%%%%%%%%%%%%%%%%%%%%%%%%%%%%%%%%%%%%%%%%%
\subsection{Influence of $\alpha$ and $\beta$ on the DW dynamics}\label{subsec:automotion}
%%%%%%%%%%%%%%%%%%%%%%%%%%%%%%%%%%%%%%%%%%%%%%%%%%%%%%%%%%%%%%%%%%%%%%%%%%%%%%%%%%%%%%%%%%%%%%%%%%%%%%%%%%%%
%%%%%%%%%%%%%%%%%%%%%%%%%%%%%%%%%%%%%%%%%%%%%%%%%%%%%%%%%%%%%%%%%%%%%%%%%%%%%%%%%%%%%%%%%%%%%%%%%%%%%%%%%%%%

%%%%%%%%%%%%%%%%%%%%%%%%%%%%%%%%%%%%%%%%%%%%%%%%%%%%%%%%%%%%%%%%%%%%%%%%%%%%%%%%%%%%%%%%%%%%%%%%%%%%%%%%%%%%
\begin{figure*}[ht]
  \includegraphics[width=18cm]{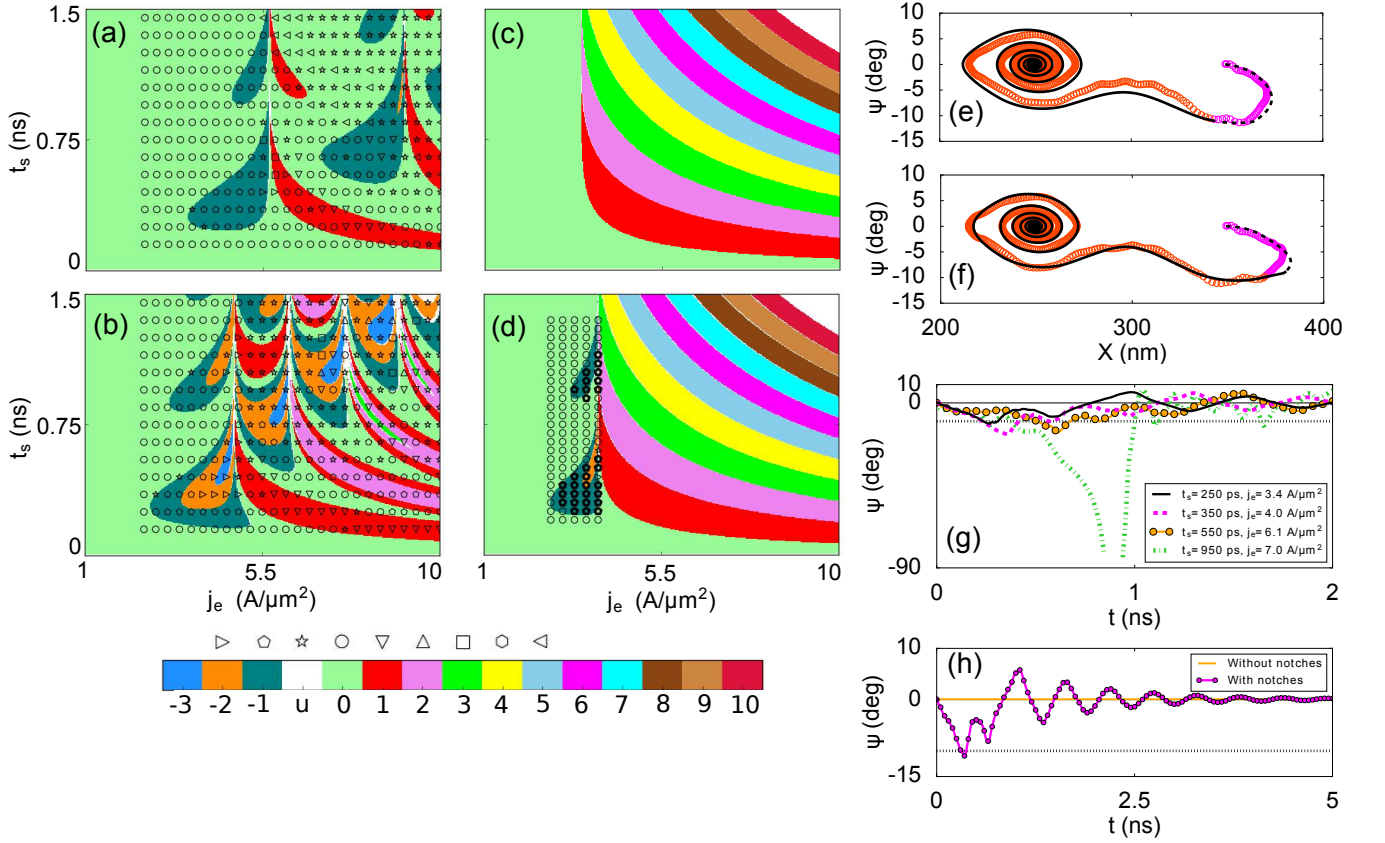}\\
 \caption{\label{Fig.5} (Color online) Influence of the damping parameter $\alpha$ and non-adiabatic parameter $\beta$ on the phase diagram for an infinite strip at T=0K: (a) $\alpha$ = 0.05 and $\beta$ = 0, (b) $\alpha$ = 0.02 and $\beta$ = 0, (c) $\alpha$ = $\beta$ = 0.05, (d) $\alpha$ = $\beta$ = 0.02. A sequence of symmetric pulses with same characteristics as in Fig.~\ref{Fig.2} are applied. The scattered symbols represent the micromagnetic results, while the colored regions are calculations with the 1D model. Trajectories in the phase space (X,$\psi$) corresponding to the -1 band in panels (b) respectively (d): (e) t$_s$ = 250 ps, j$_e$ = 3.4 A/$\mu$m$^{2}$,  $\alpha$ = 0.02, $\beta$ = 0 and (f) t$_s$ = 300 ps, j$_e$ = 3.6 A/$\mu$m$^{2}$, $\alpha$ = $\beta$ = 0.02. The scattered symbols and the full line represent the micromagnetic and the 1D results respectively. The dotted part of the full line indicates the applied pulse duration. (g) DW angle variation for $\alpha$ = 0.02, $\beta$ = 0 for several points in the panel (b). (h) Comparison of the DW angular variation for $\alpha$ = $\beta$ = 0.02 and t$_s$ = 300 ps, j$_e$ = 3.6 A/$\mu$m$^{2}$ for the infinite strip with notches (band -1 in panel (d)) and strip without notches.}
\end{figure*}
%%%%%%%%%%%%%%%%%%%%%%%%%%%%%%%%%%%%%%%%%%%%%%%%%%%%%%%%%%%%%%%%%%%%%%%%%%%%%%%%%%%%%%%%%%%%%%%%%%%%%%%%%%%%

The influence of the damping parameter $\alpha$ and of the non-adiabatic parameter $\beta$ on the phase diagram for an infinite strip at T=0K is detailed in Fig.~\ref{Fig.5}. The damping parameter $\alpha$ was varied between the 0.02 and 0.05, which correspond to the zero and room temperature values\cite{Gilmore}. The non-adiabatic parameter was varied between zero and 2$\alpha$. The case with $\beta$ = 2$\alpha$ is shown in Fig.~\ref{Fig.2}(b) and Fig.~\ref{Fig.4}(b) for $\alpha$ = 0.05 and 0.02 respectively. The computations with $\beta$ = 0 are presented in panels (a) and (b) of Fig.~\ref{Fig.5}, while the ones with $\beta$ = $\alpha$ are displayed in panels (c) and (d). We observe that for $\alpha$ = 0.05, when $\beta$ is diminished from 2$\alpha$ to zero, important changes appear in the phase diagram only when $\beta$ is less than $\alpha$. In this case (panel (a)), the results obtained with the 1D model (colored regions) exhibit only the first band (band +1) with large pockets of negative numbered bands. In all cases, the DW is pushed initially by the current pulse in the positive direction (to the right), so a negative band express a DW position at the end of the pulse to the left of the initial notch, in the opposite direction of the STT. The micromagnetic computations confirm this behavior, which was predicted and observed before\cite{Thomas1,Thiaville1,Chauleau,Vogel}. Due to the pinning potential, the DW deforms and can change its internal structure giving rise to a transient motion associated with DW inertia. The transient DW movement is proportional to the variation of the generalized angle of the wall:

%%%%%%%%%%%%%%%%%%%%%%%%%%%%%%%%%%%%%%%%%%%%%%%%%%%%%%%%%%%%%%%%%%%%%%%%%%%%%%%%%%%%%%%%%%%%%%%%%%%%%%%%%%%%
\begin{equation}\label{eq2}
\delta X = - \frac{\Delta}{\alpha} \left( 1 - \frac{\beta}{\alpha} \right) \delta \psi
\end{equation}
%%%%%%%%%%%%%%%%%%%%%%%%%%%%%%%%%%%%%%%%%%%%%%%%%%%%%%%%%%%%%%%%%%%%%%%%%%%%%%%%%%%%%%%%%%%%%%%%%%%%%%%%%%%%

\noindent with $\psi$ the azimuthal angle of the wall. For $\beta$ = 0, the transient motion is increased for large DW width or small damping parameter. This effect is displayed in Fig.~\ref{Fig.5}(b) for $\alpha$ = 0.02, as compared with panel (a), where an increased number of negative bands are visible. For the ultra-short current pulses used, which are comparable with the DW damping time $\tau_d$, the transient effects dominate at low damping parameter. 

A discrepancy is found between the micromagnetic results (symbols) and the 1D results in the upper right quadrant (high current, longer pulse) for the phase diagrams with $\beta$ = 0. This discrepancy is due to the large angle variation at large current and longer pulse, that leads to the transformation of the transverse DW and the creation of an antivortex close to the initial notches before depinning\cite{YuanEPJB} (see figure and movie in Ref.\onlinecite{SM}). DW velocity boosting was predicted through antivortex generation at a singular notch in a nanostrip\cite{Yuan}. The antivortex disappear quickly after the end of the current pulse, and in certain cases can reverse the orientation of the magnetization at the center of the transverse DW leading to DW motion in opposite direction. The antivortex does not appear in all the computed micromagnetic results in the upper right quadrant of these phase diagrams. In the 1D model, the antivortex nucleation is not taken into account.    

As is obvious from Eq.\eqref{eq2}, for $\beta$ = $\alpha$ the transient motion is somehow blocked and the DW travels rigidly. This seems to be the case for $\beta$ = $\alpha$ = 0.05, as computed in the phase diagram shown in Fig.~\ref{Fig.5}(c). However, when $\beta$ = $\alpha$ = 0.02, the negative bands are still present at low currents (panel (d)) without the apparition of an antivortex. This effect was verified micromagnetically on a number of points (empty symbols), which compare very well with the 1D results. The pocket form of the -1 band is respected with a small shift, even a -2 band was observed micromagnetically corresponding to the small -2 pocket inside the -1 band. The DW motion in the case $\beta$ = $\alpha$ can be well explained analyzing the 1D equations of motion of the DW:     

%%%%%%%%%%%%%%%%%%%%%%%%%%%%%%%%%%%%%%%%%%%%%%%%%%%%%%%%%%%%%%%%%%%%%%%%%%%%%%%%%%%%%%%%%%%%%%%%%%%%%%%%%%%%
\begin{align}
\label{eq3}
(1+\alpha^2)\dot{X} =& -\frac{\alpha\gamma\Delta}{2\mu_0M_s S}\frac{\partial E}{\partial X} + \frac{\gamma\Delta}{2}H_k\sin 2\psi    + (1+\alpha\beta)u \nonumber\\ 
(1+\alpha^2)\dot{\psi} =& -\frac{\gamma}{2\mu_0M_s S}\frac{\partial E}{\partial X} -\frac{\gamma\alpha}{2}H_k\sin 2\psi  + \frac{\beta-\alpha}{\Delta}u
\end{align}
%%%%%%%%%%%%%%%%%%%%%%%%%%%%%%%%%%%%%%%%%%%%%%%%%%%%%%%%%%%%%%%%%%%%%%%%%%%%%%%%%%%%%%%%%%%%%%%%%%%%%%%%%%%%

\noindent where S is the section of the wire, $\gamma$ the gyromagnetic ratio, and $H_k$ the DW demagnetizing field. $E$ is the pinning potential energy which is assumed parabolic inside the notch. The DW width variation is given by $\Delta(t) = \Delta[\Psi(t)]=\pi\sqrt{\frac{2A}{\mu_{0}{M_{S}}^{2}\sin^{2}{\psi}+\mu_{0}M_{S}H_{k}}}$. From the above equations, one can notice that the azimuthal DW angle $\psi$ and DW position $X$ depend on the pinning potential (restoring force) created by the symmetric notches. When a current pulse is applied, the DW is first compressed and distorts on the potential barrier moving in the direction of the STT, while the azimuthal angle decreases in some cases below -10$^{\circ}$ (dotted line), as is displayed in Fig.\ref{Fig.5}(h) (videos and additional figures in Ref.\onlinecite{SM}). Initially, the STT pushes the DW in the positive direction, resulting in a positive DW velocity and a negative restoring force (negative DW angle). If the DW does not have enough velocity to surpass the potential barrier in this direction, it goes down the potential well towards the center of the notch and the restoring force still stays negative, while the velocity and angle continue to decrease with the velocity becoming negative. When the DW starts to mount the potential well in the other direction, the restoring force becomes positive so the velocity and DW angle start to increase (Fig.~\ref{Fig.5}(f)). When the current pulse ends, the velocity decreases abruptly (increases in absolute value) and $\psi$ increases lightly do to the short fall time. At this moment, the DW position inside the potential well is important as it imposes a restoring force in the direction of movement or in the opposite direction. However, the velocity and the DW angle are equally important for the jump to the previous notches to happen\cite{SM}. The jump occurs therefore due to a combination of factors, even if the restoring force is positive or negative. The presence of the notch pinning potential induces the DW distortion and therefore the transient motion, as for the same pulse parameters no observable angle variation is determined for a perfect strip. As the variation of the DW angle is directly related to the variation in position, the automotion is possible in this case triggered by the pinning induced DW distortion, restoring potential force and smallness of the damping parameter.

In Fig.~\ref{Fig.5}(g), several angular variations are presented corresponding to the -1 and -2 bands from panel (b). We observe that the amplitude of angular variation is directly proportional with the spatial displacement. When an antivortex appears, the DW angle rotates out of plane and the DW position does not correspond anymore to the 1D results and the motion opposite to the STT can be completely blocked. In general, we observe the automotion in all the cases when the DW angle increases above 10$^{\circ}$ (in absolute values) during or after the current pulse and the maximum DW velocity is close or superior to 400m/s (details in Ref.\onlinecite{SM}). Exactly at the boundary between the -1 band and the zero band (pinned state), a small increase in the azimuthal angle of 0.2$^{\circ}$ and of the DW velocity by 5m/s at the end of the current pulse, between two points in different bands, is enough to promote a DW jump to a previous notch.

The transient effects also appear when $\beta > \alpha$, as depicted in Fig.~\ref{Fig.4} for $\beta = 2\alpha$ = 0.04. A low value of the damping parameter $\alpha$ is required to obtain observable consequences. A particularity of the case $\beta > \alpha$ is that the transient effects oppose or amplify the STT, as negative bands are determined and a re-entrant transition is seen at higher currents and pulse length in panels (b) and (d). The transient effects depend on the pulse shape\cite{SM}, as for the asymmetric pulse with t$_r$ = 5 ps, t$_f$ = 300 ps, the band -1 is barely visible and no re-entrant transition of the bands is seen, while the -1 band increases when the pulse is symmetric and continues to increase, with even a second -1 band appearing at larger t$_s$ for the asymmetric pulse with t$_r$ = 300 ps, t$_f$ = 5 ps. The fall time $t_f$ plays an important role in the value of the DW velocity and DW angle at the end of the current pulse, as a short $t_f$ leads to a very high DW velocity and a higher DW angle at the end of the pulse increasing the impact of transient effects and inducing a DW depinning. Contrary to earlier beliefs\cite{Thomas1}, the DW depinning does not necessarily result from a large DW angle or the DW position at the end of the pulse. A maximum DW angle and DW velocity large enough during the current pulse suffice to ensure for example the jump to the previous notch, even if the angle is not that large (and at its maximum) at the pulse end, as observed in the case of the asymmetric pulse with t$_r$ = 5 ps, t$_f$ = 300 ps (Ref.\onlinecite{SM}). This is purely an transient effect due to a combined action of the pinning potential, low $\alpha$ and pulse shape.

\begin{figure}[ht]
  \includegraphics[width=8cm]{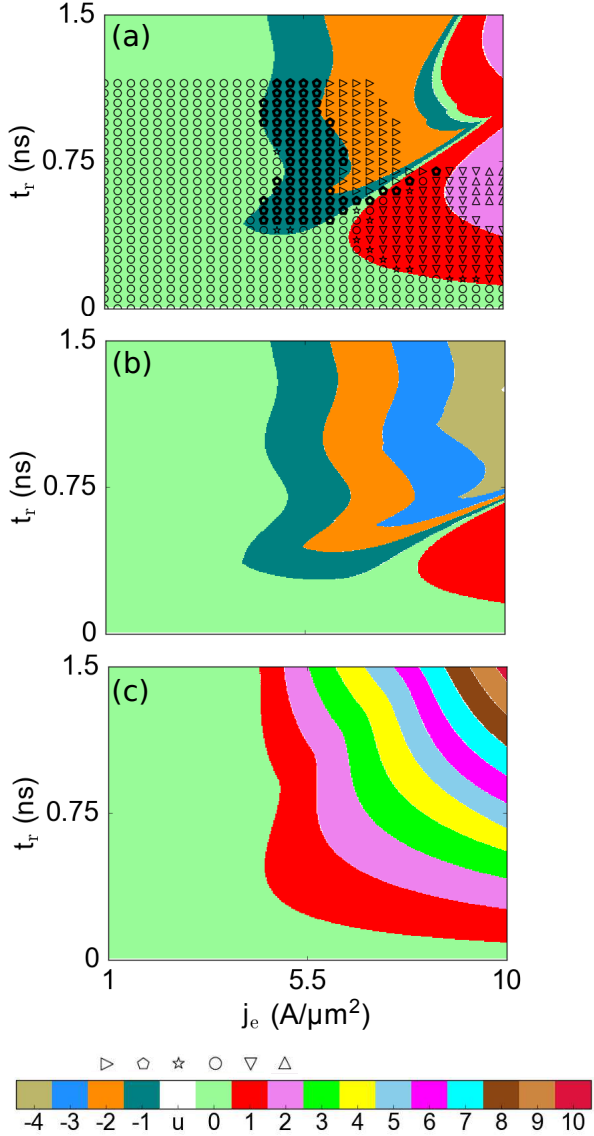}\\
 \caption{\label{Fig.6} (Color online) Influence of the pulse raise time t$_r$ on the phase diagram for a DW at T=0K in an infinite nanostrip for different parameters $\alpha$ and $\beta$. The parameter space is the raise time vs. current amplitude. In all cases,  t$_s$ = t$_f$ = 5 ps and t$_z$ = 10 ns. (a) $\alpha$ = 0.02 and $\beta$ = 0.04, (b) $\alpha$ = $\beta$ = 0.02, and (c) $\alpha$ = 0.05 and $\beta$ = 0.1. In (a) the micromagnetic results (scattered symbols) are compared to the 1D model (colored regions), while in (b) and (c) only 1D model results are shown. At high currents, the micromagnetic results give several unintended states (the empty scattered symbol region to the right of (a)).}
\end{figure}

The influence of the rise time t$_r$ on the phase diagrams is presented in Fig.~\ref{Fig.6} for t$_s$ = t$_f$ = 5 ps and several parameters $\alpha$ and $\beta$. The rise time is varied between 0 and 1.5 ns (case of a very asymmetric pulse). A first observation is that the first depinning current is increased as the force term that depends on the current derivative in Eq.\eqref{eq1} diminishes. The first depinning current actually oscillates with t$_r$ due to resonants effects, as the resonant frequency of the potential well is around 1.75 GHz. This resonant effect are more important for the bands with negative numbers for low $\alpha$, as is depicted in panels (a) and (b) for $\alpha$ = 0.02 and $\beta$ = 0.4 or 0.2 respectively. The 1D model gives good quantitative results as compared with the micromagnetic calculations, as shown in panel (a), although a small shift is again found at high currents and more unwanted states. When $\beta \leq \alpha$, the positive bands are shifted to very high current, with only negative bands remaining for $\beta$ = 0 (image not shown). The phase diagram for $\alpha$ = 0.05 and $\beta$ = 0.1 is also shown in panel (c), that still does not presents negative bands but an the same oscillation of the first depinning current and of the interband transitions. 

The influence on the phase diagram is less dramatic when varying the fall time t$_f$ (images shown in Ref.\onlinecite{SM}). For $\alpha$ = 0.02 and $\beta > \alpha$ no negative bands appear and the positive bands are shifted to larger currents. When $\beta \leq \alpha = 0.02$, only a small pocket of the -1 band appears and the positive bands are shifted as compared with Fig.~\ref{Fig.5}(d).

%%%%%%%%%%%%%%%%%%%%%%%%%%%%%%%%%%%%%%%%%%%%%% FIGURE 6 %%%%%%%%%%%%%%%%%%%%%%%%%%%%%%%%%%%%%%%%%%%%%%%%%%%%%%%%%%%%%%
%%%%%%%%%%%%%%%%%%%%%%%%%%%%%%%%%%%%%%%%%%%%%%%%%%%%%%%%%%%%%%%%%%%%%%%%%%%%%%%%%%%%%%%%%%%%%%%%%%%%%%%%%%%%
%%%%%%%%%%%%%%%%%%%%%%%%%%%%%%%%%%%%%%%%%%%%%%%%%%%%%%%%%%%%%%%%%%%%%%%%%%%%%%%%%%%%%%%%%%%%%%%%%%%%%%%%%%%%%%%%%%%%%%%%%%%%%%%%%%%%%%%%%%%%%%%%%%%%%%%%%%%%%%%%%%%%%%%%%%%%%%%%%%%%%%%%%%%%%%%%%%%%%%%%%%%%%%%%%%%%%%%%

%%%%%%%%%%%%%%%%%%%%%%%%%%%%%%%%%%%%%%%%%%%%%%%%%%%%%%%%%%%%%% Section %%%%%%%%%%%%%%%%%%%%%%%%%%%%%%%%%%%%%%%%%%%%%%%%%%%% %%%%%%%%%%%%%%%%%%%%%%%%%%%%%%%%%%%%%%%%%%%%%%%%%%%%%%%%%%%%%%%%%%%%%%%%%%%%%%%%%%%%%%%%%%%%%%%%%%%%%%%%%%%%%%%%%%%%%%%%%%%%%%%%%%%%%%%%%%%%%%%%%%
\section{Discussion and conclusion}\label{Discussion}
%%%%%%%%%%%%%%%%%%%%%%%%%%%%%%%%%%%%%%%%%%%%%%%%%%%%%%%%%%%%%%%%%%%%%%%%%%%%%%%%%%%%%%%%%%%%%%%%%%%%%%%%%%%%%%%%%

The presence of artificial constrictions in a nanostrip influences drastically the movement of a magnetic DW. When ultra-short current pulses are applied, the DW can exhibit an important distortion at the notch depending on the pulse characteristics. Thereby, the DW displays inertia-like effects, which can have dramatic consequences on its transient displacement. These effects depend largely on the damping parameter $\alpha$ and on the non-adiabatic parameter $\beta$. For $\beta < \alpha$, these effects generally oppose the STT effect after the pulse end and DW motion in the direction of the electric current is possible. If $\beta > \alpha$, these effects oppose or amplify the STT effect and jumps to the left or the right notches are possible after the end of the pulse. The transient effects in this case depend on the pulse characteristics. This could constitute another way of experimentally compare the two parameters $\alpha$ and $\beta$ for a ferromagnetic material.

At room temperature, the jump probability to the desired notches decreases with increasing band number, each band number corresponding to a positioning to desired notches in the direction of the STT. Maximum positioning probability is reached only for very short rise and fall time to the nearest neighbors notches only. Therefore, to shift reliably the DWs between notches, current pulses corresponding to displacement from one symmetric notch to the closest neighbors should be used. The shape of the current pulse influences the depinning current and shifts the bands. The phase diagram for the case of two domain walls situated at different symmetric notches (images not shown), that are displaced by the same current pulse in the same direction, is very similar with the ones presented in Sec.\ref{Results}, but the bands are narrower and the interband depinning is larger between the first bands. 

The main drawback of the classical DW displacement under STT alone compared with more exotic torques (of spin-orbit origin) is the high current values necessary. The current induces Joule heating in the nanostrip, that can largely increase the temperature and could even destroy the ferromagnetic state. The increase in temperature is even more important at the constrictions in the nanostrip. Several theoretical studies were dedicated to Joule heating in nanowires\cite{You,Fangohr,Moretti,Lepadatu}, usually considering a standard Py nanostrip on a Si/SiO$_2$ substrate. We evaluated the temperature increase for our Ni strips on different substrates like pure Si, SiO$_2$ or Ni$_3$Si$_4$ membranes for a current pulse length of 1ns. On pure Si, considering an infinite 3D substrate\cite{You}, the temperature increase is negligible being of 4K for j = 5 A/$\mu$m$^2$ (17K for j = 10 A/$\mu$m$^2$). However, on SiO$_2$ substrate of 300nm thickness\cite{Fangohr}, the temperature increase is larger being 26K for j = 5 A/$\mu$m$^2$ (103K for j = 10 A/$\mu$m$^2$). If 100nm thick Ni$_3$Si$_4$ membranes are used\cite{Bocklage,Fangohr}, the temperature increase is of 41K for j = 5 A/$\mu$m$^2$ (163K for j = 10 A/$\mu$m$^2$). The variation in temperature depends on the material parameters of the substrate (like thermal conductivity) and on the conductivity of the nanostrip. In the above estimations, the bulk Ni room temperature conductivity was used ($\sigma^{-1}$ = 7.3 $\mu\Omega\cdot$cm)\cite{White}. If the dimensions are reduced to nanometers\cite{Ou}, the conductivity of Ni can vary drastically (a factor four) and the temperature increase can be more important, but in the main time the length of the current pulse that amounts to the maximum probability at room temperature is around 0.1ns which reduces significantly the temperature increase. 

In conclusion, systematic DW motion between precise artificial pinning constriction by very short current pulses is possible at room temperature in a classical ferromagnet. The constrictions induce DW distortions and important transient effects can be observed. Depending on the ratio $\beta/\alpha$, the inertia-like effect can oppose or amplify the STT effect on the DW motion after the pulse end. As the value of $\beta$ is still under debate, this could constitute another way of determining its relative value. Our results open the path to DW motion in both directions by unipolar current pulses.

%%%%%%%%%%%%%%%%%%%%%%%%%%%%%%%%%%%%%%%%%%%%%%%%%%%%%%%%%%%%%%%%%%%%%%%%%%%%%%%%%%%%%%%%%%%%%%%%%%%%%%%%%%%%%%%%%%%%%%%%%%%%%%%%%%%%%%

\end{document}